\documentclass[12pt]{article} 
\usepackage{multibib}
\usepackage[labelfont=bf]{caption}
\usepackage{booktabs}
\usepackage{graphicx}
\usepackage{authblk}
\usepackage[utf8]{inputenc}
\usepackage[a4paper, margin=1in]{geometry} 
\usepackage{caption}
\usepackage[version=4]{mhchem}
\usepackage{amsmath, amssymb, mathtools} 
\usepackage{physics}
\usepackage{chemformula}
 
\usepackage{chemfig}
\usepackage{graphicx} 
\usepackage{float} 
\usepackage{subcaption} 
\usepackage{authblk} 
\usepackage{hyperref} 
\hypersetup{
    colorlinks=true,
    urlcolor=tiel,
    linkcolor=orange,
    citecolor=olive
}
\usepackage{cite}

\usepackage{outlines}  
\usepackage{array}

\title{Magnetic Field-dependent Isotope Effect Supports Radical Pair Mechanism in Tubulin Polymerization}

\author[1]{Hadi Zadeh-Haghighi\textsuperscript{*†}}
\author[1]{Caleb R. Siguenza\textsuperscript{*}}
\author[2]{Robert P. Smith}
\author[3]{Christoph Simon}
\author[1]{Travis J.A. Craddock\textsuperscript{‡}}

\affil[1]{Department of Biology, Waterloo Institute for Nanotechnology, University of Waterloo, Waterloo, ON, Canada N2L 3G1}
\affil[2]{Department of Medical Education, Cell Therapy Institute, Kiran Patel College of Allopathic Medicine, Nova Southeastern University, Fort-Lauderdale-Davie, FL, USA 33328}
\affil[3]{Department of Physics and Astronomy, Institute for Quantum Science and Technology, Hotchkiss Brain Institute, University of Calgary, Calgary, AB, Canada T2N 1N4}

\date{}

\begin{document}

\maketitle
\renewcommand{\thefootnote}{\fnsymbol{footnote}}

\footnotetext[1]{These authors contributed equally to this work.}
\footnotetext[2]{hadi.zadehhaghighi@uwaterloo.ca}
\footnotetext[3]{travis.craddock@uwaterloo.ca}

\begin{abstract}

Weak magnetic fields and isotopes have been shown to influence biological processes; however, the underlying mechanisms remain poorly understood, particularly because the corresponding interaction energies are far below thermal energies, making classical explanations challenging or impossible. Microtubules, dynamic cytoskeletal fibers, offer an ideal system to test weak magnetic field effects due to their self-assembling capabilities, sensitivity to magnetic fields, and their central role in cellular processes. In this study, we use a combination of experiments and simulations to explore how nuclear spin dynamics affect microtubule polymerization by examining interactions between magnesium isotope substitution and weak magnetic fields. Our experiments reveal an isotope-dependent effect, which can be explained via a radical pair mechanism, explicitly arising from nuclear spin properties rather than isotopic mass differences. This nuclear spin-driven isotope effect is notably enhanced under an applied weak magnetic field of approximately 3 mT. Our theoretical model based on radical pairs achieves quantitative agreement with our experimental observations. These results establish a direct connection between quantum spin dynamics and microtubule assembly, providing new insights into how weak magnetic fields influence cellular and biomolecular functions.

\end{abstract}
\textbf{Keywords:} quantum biology, radical pair mechanism, magnetic field effects, magnetic isotope effects, microtubule cytoskeleton, magnesium

\section*{Introduction}

Weak magnetic fields and isotope effects have been observed to influence biological systems, yet the underlying mechanisms remain debated. Biological processes are generally understood through biochemical and thermodynamic principles; however, some reactions demonstrate sensitivities that classical models cannot easily explain—especially given that the corresponding energies from such weak magnetic fields are far below thermal energies at physiological temperatures. Notably, these effects have been reported across diverse systems, including electron transfer in cryptochrome \cite{XuCRY2021,Bradlaugh_2023}, stem cells \cite{VanHuizen2019}, neurogenesis \cite{Zhang2021}, axon growth \cite{Dufor2019}, the circadian clock \cite{Yoshii2009}, and cellular autofluorescence \cite{Ikeya2021}. These observations suggest that certain biochemical processes may depend on quantum properties such as electron and nuclear spin dynamics \cite{ZadehHaghighi2022}. Consequently, rigorous experimental validation of quantum mechanical mechanisms in biologically relevant contexts remains essential.

\begin{figure}[ht]
    \centering
    \includegraphics[width=0.45\textwidth]{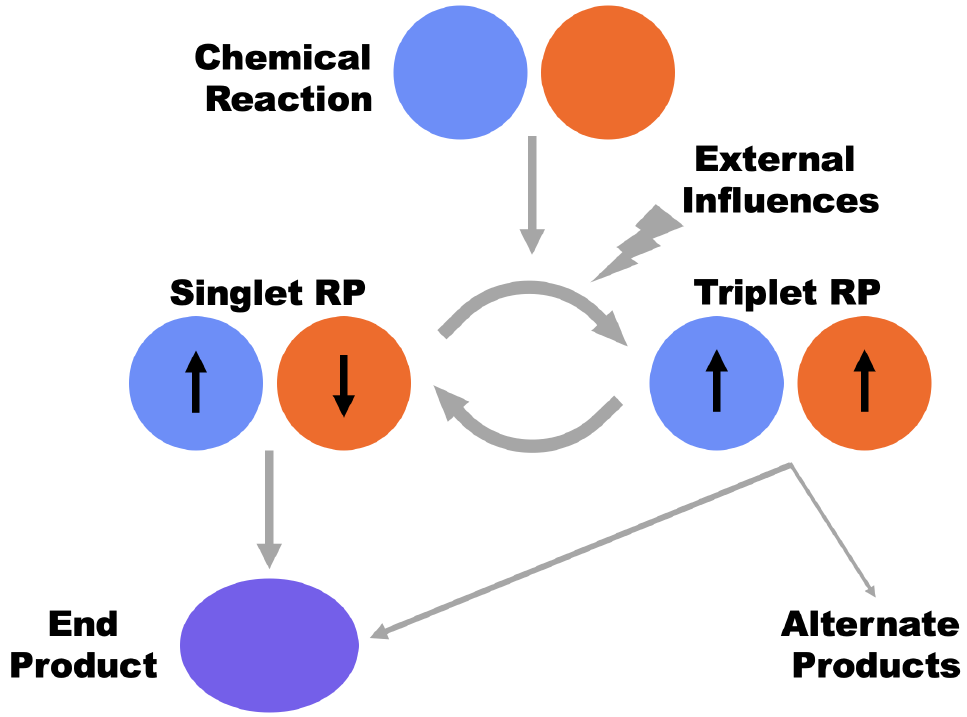}
    
    \caption{\scriptsize\textbf{Radical Pair Mechanism } A transient radical pair (RP) can be formed during a chemical reaction which may interconvert between singlet and triplet states (arrows indicate electron spin) depending on external influences such as magnetic fields or nearby nuclear spins. End products may be formed from the singlet or triplet state at different rates, while the triplet state may also lead to alternate chemical products.}   
    \label{fig:figure1}
\end{figure}

A leading explanation for these effects is the radical pair mechanism (RPM)  \cite{ZadehHaghighi2022, Steiner1989, Hore2016}. The RPM involves transient pairs of atoms or molecules each with an unpaired electron (i.e., radicals) that are formed during chemical reactions (Figure \ref{fig:figure1}A). These radical pairs can exist in two distinct quantum states: a singlet state, where the electron spins are entangled and their combined spin is zero, and a triplet state, where the spins are aligned in a way that gives a total spin of one. Due to the interaction of the electron spins with the environment, external magnetic fields, and nuclear spins, these radical pairs can oscillate between singlet and triplet spin states \cite{Steiner1989,Timmel1998}. The rate at which a radical pair recombines to complete a chemical reaction depends on its spin state. This spin-dependent recombination underlies magnetically sensitive spin chemistry, explaining how weak magnetic fields influence biochemical processes \cite{Schulten1976,Hore2020}. Isotopes of elements such as lithium, magnesium, calcium, or zinc can influence this process through hyperfine coupling, which involves interactions between their nuclear spins and electron spins. This coupling establishes a clear connection between spin-dependent isotope effects and the sensitivity of biological systems to weak magnetic fields \cite{ZadehHaghighi2022}. 

In a biological context, flavin-tryptophan radical pairs in cryptochrome have been shown to generate magnetic field effects \cite{XuCRY2021}. Beyond this specific case, the RPM can provide a general framework for explaining weak magnetic field effects across diverse biological systems. However, many of these systems lack cryptochrome or even flavin, and some also exhibit isotope effects that remain poorly understood \cite{ZadehHaghighi2022}. Testing whether RPM-based mechanisms are at play in these contexts requires carefully designed experiments in other established molecular systems. Observation of combined weak field and isotope effects consistent with theoretical predictions would be a strong indication of radical pair involvement in biology. To our knowledge, no experiment has been able to conclusively show both weak magnetic field and isotope effects in a biologically relevant chemical system that is consistent with a theoretical description of radical pair dynamics.  That is the objective of this investigation.

Microtubules offer an ideal biochemical system for assessing the broader applicability of the RPM beyond well-studied systems like bird navigation and cryptochrome. Microtubules are dynamic protein polymers essential for cell structure, intracellular transport, and cellular organization \cite{Brouhard2018,Akhmanova2015,Kuo2021} in all eukaryotic systems. Their rapid polymerization and depolymerization enable cytoskeletal reorganization during processes like cell division and neuronal growth \cite{Gudimchuk2021,Leterrier2021,Akhmanova2022}, making them targets for treating cancer and neurodegenerative diseases \cite{Gudimchuk2021}. Known factors that regulate microtubule polymerization dynamics include magnesium (Mg), pH and temperature, as well as removal of the weak geomagnetic field (GMF) (25-65 $\mu$T) \cite{Wang2008}. How small changes in magnetic fields influence microtubule polymerization dynamics remains unknown. However, a recent theoretical investigation has proposed that spin dynamics can directly influence microtubule density through effects on polymerization rates \cite{ZadehHaghighi2022MT}. This RPM-based model has theoretically reproduced the experimentally observed hypomagnetic field (HMF) ($<$ 5 $\mu$T) effects on microtubule density \cite{Wang2008}, and further predicts isotope-dependent and applied weak magnetic field effects \cite{ZadehHaghighi2022MT}.

In this study, we investigate the influence of weak magnetic fields and different Mg isotopes on microtubule polymerization dynamics. We measure changes in optical density to assess microtubule polymerization in the presence of Mg isotopes with and without nuclear spins in the absence or presence of a weak magnetic field while accounting for confounding factors such as pH, temperature, and differences in isotope weight. We report experimental evidence demonstrating substantial changes in microtubule polymerization dynamics induced by \textsuperscript{25}Mg in the presence of a weak external magnetic field. We further show that a general RPM model is capable of reproducing experimentally observed Mg isotope and weak magnetic field effects on microtubule polymerization.

\section*{Results}
\subsection*{Weak Magnetic Field Enhances Tubulin Polymerization in the Presence of the $^{25}$Mg Isotope}

\begin{figure}
    \centering
    \includegraphics[width=0.9\textwidth]{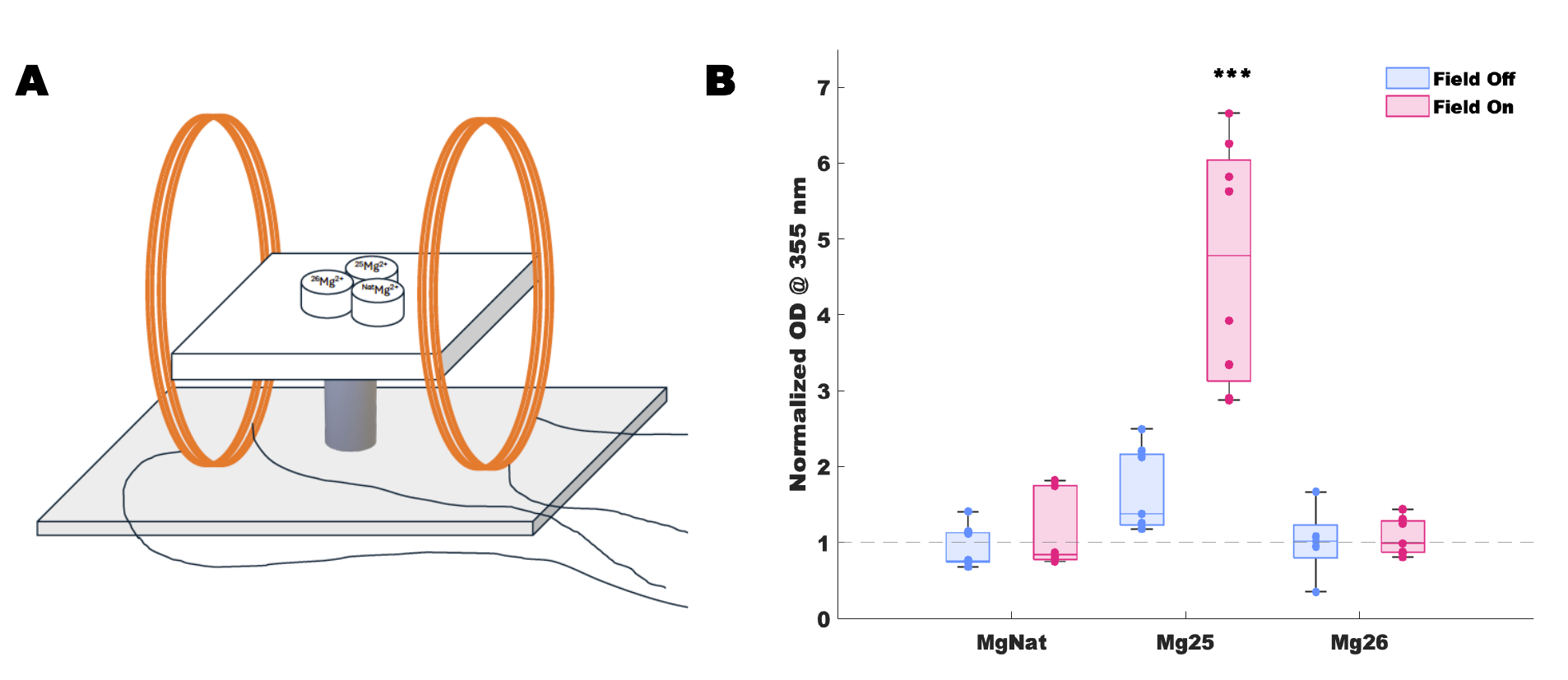} 
    \caption{\scriptsize\textbf{Isotope Effect on Tubulin Polymerization in the Absence and Presence of a 3 mT Magnetic Field.} (A) Schematic representation of the magnetic field setup using Helmholtz coils. Tubulin samples containing $^{Nat}$Mg(79\% $^{24}$Mg, 10\% $^{25}$Mg, 11\% $^{26}$Mg), $^{25}$Mg, and $^{26}$Mg were placed at the center of the Helmholtz coils which generated a uniform magnetic field of $2.99 \pm 0.02$ mT (Field On). (B) Box plots showing the median line, upper and lower quartile box, maximum and minimum whiskers, and scatter of individual replicates for the final optical density (OD) measure of tubulin polymerization for $^{Nat}$Mg, $^{25}$Mg, and $^{26}$Mg in the absence (blue) and presence (magenta) of the applied 2.99 \(\pm\) 0.02 mT magnetic field. OD values are in arbitrary units relative to the final values of tubulin polymerization with $^{26}$Mg in the absence of an applied magnetic field. *** $p < 10^{-7}$, difference from $^{25}$Mg in the absence of an applied field as determined by a two-way ANOVA with a Tukey-Kramer posthoc test. Notably, neither \(^{\text{Nat}}\text{Mg}\) nor \(^{26}\text{Mg}\) show a significant difference between field-on and field-off conditions.}
    \label{fig:figure2}
\end{figure}

To investigate the role of nuclear spin in microtubule polymerization we measured the effect of Mg isotopes on tubulin assembly dynamics, both in the presence and absence of a uniform 2.99$\pm$0.02 mT magnetic field applied using a Helmholtz coil (Figure \ref{fig:figure2}A). In-house made polymerization buffers were all made with identical pH and shown to reproduce standard manufacturer polymerization curves (Figure \ref{fig:figureS1}). The high enrichment values of the isotopes ensured minimal presence of paramagnetic ions (cobalt, copper, iron, manganese, molybdenum, vandium etc.) at concentrations less than 0.6 $\mu$M (Figure \ref{fig:figureS2} and Figure \ref{fig:figureS3}). Analysis of tubulin polymerization as measured by optical density (OD) at 355 nm indicated that there was a significant enhancement in total tubulin polymerized ($p<10^{-7}$) in the presence of $^{25}$Mg compared to $^{Nat}$Mg and $^{26}$Mg with the application of the magnetic field (Figure \ref{fig:figure2}B).

To directly test alternative hypotheses that could explain the effects observed in Figure \ref{fig:figure2}, we specifically examined whether differences in isotope mass, well-to-well temperature differences, or well-to-well magnetic field differences could account for the observed results (Figure \ref{fig:figureS5}B-C). No differences were observed to be due to increasing isotope mass (i.e., kinetic isotope effects) as $^{25}$Mg is both lighter than $^{26}$Mg and heavier than $^{Nat}$Mg which is primarily composed of $^{24}$Mg (79\%).  Analysis of well-to-well temperature variations did not show a significant correlation between OD and temperature for any condition ($p> 0.13$) and did not show a significant contribution to the observed effect ($p> 0.26$)  (Figure \ref{fig:figureS4}A). Analysis of well-to-well magnetic field variations also did not show a significant correlation between OD and magnetic field for either the field off ($p>0.11$) (Figure \ref{fig:figureS4}B) and field on conditions ($p> 0.19$) (Figure \ref{fig:figureS4}C). All data, including raw and normalized polymerization data, temperature and magnetic field measures, and statistical analyses, are provided in the Supplementary Files.

\subsection*{A Radical Pair Model of Magnetic Field and Isotope Effects on Microtubule Dynamics}
To explain our experimental observations, we employed a simple ordinary differential equation model of microtubule dynamics \cite{Craddock2012} describing the dynamic exchange between free tubulin molecules ($Tu$) and tubulin polymerized into microtubules ($MT$) such that the time, $t$, dependent concentration of $MT$ is defined as:

\begin{equation}
[MT(t)]= \frac{k_p [P]}{k_p+k_d} (1-e^{-(k_p+k_d)t}),
\label{eq:MTC}
\end{equation}

\noindent where $k_p$ is the rate of polymerization, $k_d$ is the rate of depolymerization, and $[P]$ is the concentration of total tubulin protein (i.e., $[P]=[MT]+[Tu]$). As Mg is required for tubulin to hydrolyze GTP, and removal of the GTP cap increases microtubule depolymerization, we incorporate Mg-dependent magnetic field and isotope effects into the model by modulating $k_d$ through changes in the radical pair triplet yield as follows::
\begin{equation}
k_d' \propto \frac{\Phi_T'}{\Phi_T}k_d, 
\label{eq:ratio}
\end{equation}
where, $k_d'$, $\Phi_T$, and $\Phi'_T$ represent the modified depolymerization rate constant, the triplet yield under the standard ambient magnetic field (i.e., GMF) with a zero spin Mg isotope, and the triplet yield arising from an applied magnetic field and hyperfine interactions due to a non-zero isotope nuclear spin, respectively.  The triplet yield depends on the interaction terms, their spin relaxation rates, and the reaction rates for the radicals in the singlet and triplet states. Complete details concerning the model terms and parameters are in the Methods section. 

To incorporate the magnetic field and Mg isotope effects, we use a generic RP model. In this model, each radical (A and B) is coupled to a spin-1/2 nucleus with hyperfine coupling constants (HFCC) of $a_{A/B}$, by additionally including $^{25}$Mg coupling to electron B with HFCC of $a_{25}$. Electron A/B is subject to a spin relaxation rate $r_{A/B}$. The singlet reaction rate is $k_S$ and the triplet reaction rate is $k_T$.

Here, we focus on Zeeman and hyperfine interactions, which are not distance-dependent. In our model, we consider the case that spin correlations arise from encounters between previously uncorrelated radicals A and B (i.e., an F-pair formation mechanism) \cite{Kattnig2021}. Using the Differential Evolution algorithm \cite{Rocca2011}, the optimization of parameter values was carried out considering the following parameter bounds: \(r_A\) and \(r_B \in [10^5, 10^7]\) 1/s, \(k_S\) and \(k_T \in [10^4, 10^8]\) 1/s, \(a_A\) and \(a_B \in [-5, 5]\) mT, and \(a_{25} \in [-20, 20]\) mT. 

\begin{figure}
    \centering    \includegraphics[width=1\textwidth]{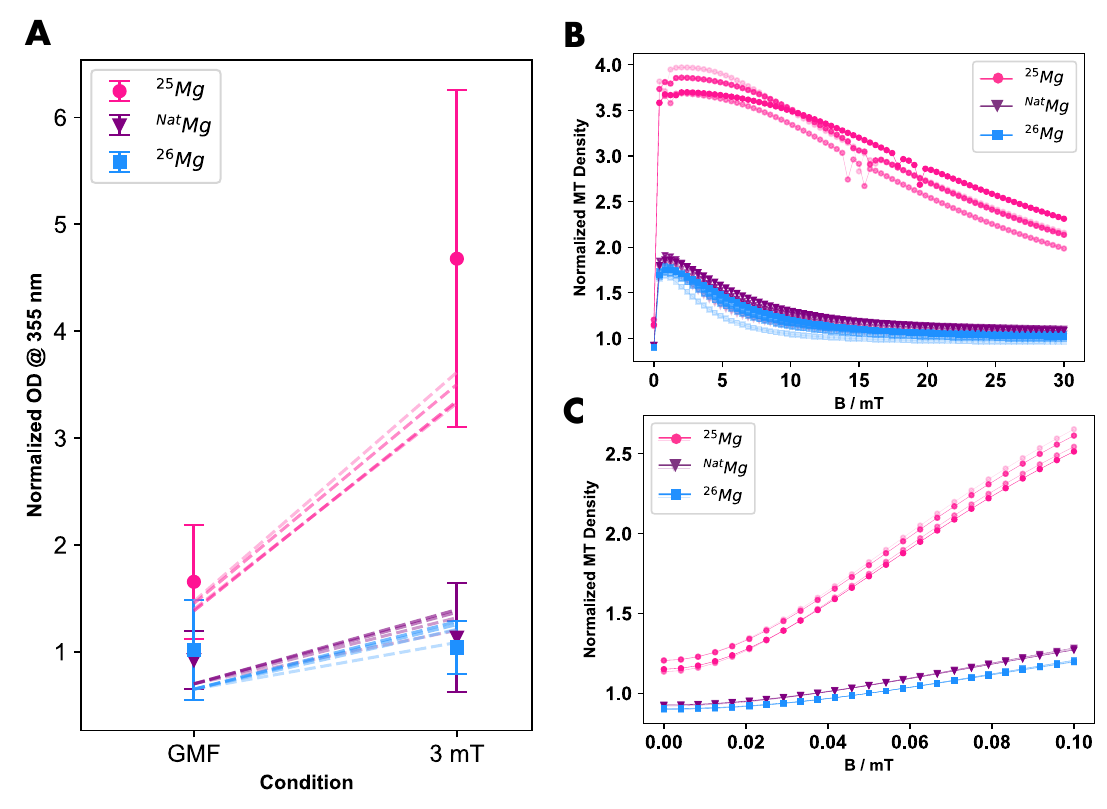} 
\caption{\scriptsize\textbf{Modeling Magnetic Field and Isotope Effects on Microtubule Dynamics.} (A) Comparison between the experimental data and radical F-pair model for microtubule density using $^{25}$Mg (magenta), $^{26}$Mg (blue) and $^{26}$Mg (purple) under two conditions with exposure of 3 mT and GMF of 0.05 mT. Experimental data are presented as mean \(\pm\) standard deviation. The dashed lines present the model. The optimization procedure yielded a diverse set of parameter values that effectively capture the observed behavior across multiple conditions. The optimized values for the recombination rates were found to range from \(k_{rA} = 1.1 \times 10^5\) to \(1.4\times 10^5~\text{s}^{-1}\), and \(k_{rB} = 1.7 \times 10^5\) to \(4.9 \times 10^5~\text{s}^{-1}\). The singlet and triplet reaction rates varied between \(k_S = 8.3 \times 10^7\) and \(9.2 \times 10^7~\text{s}^{-1}\), and \(k_T = 6.6 \times 10^5\) to \(8 \times 10^5~\text{s}^{-1}\), respectively. The HFCC were optimized as \(a_A\)=0.01 mT, \(a_{25}\) from 5.3 to 6.3~mT, and \(a_B\) from 1.4 to 2.6~mT. This range of parameters are used in (B) and (C) as well. (B) Microtubule polymerization will be further reduced with \(^{25}\text{Mg}\) at magnetic field strengths above 5 mT, compared to the levels observed at 3 mT. (C) Normalized microtubule density decreased under hypomagnetic field conditions ($<$ 5 $\mu$T). In the classical microtubule model, the polymerization rate is \(k_p = 5 \times 10^{-5}\) s\(^{-1}\), the depolymerization rate is \(k_d = 3 \times 10^{-3}\) s\(^{-1}\), and the initial tubulin concentration is 3 g/L.}
\label{fig:figure3}
\end{figure}

Experimental results show that using \(^{25}\mathrm{Mg}\) (nuclear spin \(I = 5/2\)) with the applied 3~mT field significantly increases microtubule density compared to \(^{25}\mathrm{Mg}\) at the GMF, and zero nuclear spin isotopes (\(I = 0\) effectively for \(^{Nat}\mathrm{Mg}\) and \(^{26}\mathrm{Mg}\)) at either 3~mT or GMF. In contrast, no significant change in microtubule density is observed between the zero spin isotopes for the 3~mT field and GMF conditions, indicating that the model must also account for these cases where the nuclear spin and magnetic field variations do not influence the outcome significantly. To model this substantial increase, we compare two cases: Mg ions with nuclear spin \(I = 5/2\) and those with \(I = 0\). Using Equations~\ref{eq:MTC} and \ref{eq:ratio} along with optimized parameters, the model successfully captures the field dependence of microtubule polymerization (Figure~\ref{fig:figure3}A). The optimization process produced a varied array of parameter values that successfully reflect the observed behavior under different conditions. The recombination rates were optimized to span from \(k_{rA} = 10^5\) to \(4.1 \times 10^5~\text{s}^{-1}\) and \(k_{rB} = 1.32 \times 10^5\) to \(2.6 \times 10^6~\text{s}^{-1}\). The reaction rates for singlet and triplet states ranged from \(k_S = 6.1 \times 10^7\) to \(9.4 \times 10^7~\text{s}^{-1}\) and \(k_T = 1.9 \times 10^5\) to \(9.3 \times 10^5~\text{s}^{-1}\), respectively. The HFCC were fine-tuned with \(a_A\) varying between 0 and 0.03~mT, \(a_{25}\) between 1.8 and 6.6~mT, and \(a_B\) between 0.9 and 4.2~mT (see Figure~\ref{fig:figure3}A). The corresponding data file is available in the Supplementary Files.

This model not only aligns with our experimental results but also offers predictive power for future studies. Specifically, the model predicts that microtubule polymerization will be further reduced at magnetic field strengths above 5 mT compared to the levels observed at 3 mT (see Figure \ref{fig:figure3}B). Additionally, it indicates that polymerization significantly decreases under HMF conditions compared to GMF conditions (see Figure \ref{fig:figure3}C).

\section*{Discussion}
Our experiments reveal a Mg isotope-dependent magnetic field effect on microtubule polymerization. Having ruled out influences from pH, temperature variations, the kinetic isotope effect, well-to-well magnetic field variations, and paramagnetic impurities, we conclude that the observed effect arises from the applied weak magnetic field and the non-zero nuclear spin of $^{25}$Mg (Figure \ref{fig:figure2}). These results align with an RPM description of isotope and weak magnetic field effects, strongly indicating that the RPM modulates biochemical reaction rates during microtubule polymerization \cite{ZadehHaghighi2022MT,ZadehHaghighi2023}. 

While our experiments indicate a role for Mg in the observed results, the RPM model presented here is general in nature and does not explicitly identify the radicals involved. Rather, it incorporates isotope-specific effects by including two generic spin-$\tfrac{1}{2}$ nuclear spins, one associated with each radical, as well as an additional $^{25}\mathrm{Mg}$ nuclear spin coupled to one of them. Parameters for this generic model were optimized in an unbiased way using the Differential Evolution algorithm to best capture the observed behaviors (Figure \ref{fig:figure3}).  While general in nature the parameters identified can provide insights on the potential radical pairs involved. First, the initial spin state of a radical pair can provide information on the radicals involved.  As the radicals are not predefined in our model we presented modeling results based on a fully mixed initial state. Both purely triplet and singlet initial states for the RPs in our analysis were also explored, with only triplet initial states being able to reproduce the experimental results (see Supplementary Files). Parameter optimization within our generic radical pair model further indicated that the unpaired electron on the radical opposite the Mg-bound species must be coupled to its associated nuclear spin with a very weak hyperfine coupling constant (Figure~\ref{fig:figure3}). 

All interactions involved in tubulin self-assembly are non-covalent molecular interactions except for the Mg-dependent biochemical hydrolysis of GTP. As such, the chemical intermediates of this process are the ideal candidates to be involved in a RPM. While the exact chemical mechanism of GTP hydrolysis remains actively under investigation \cite{calixto2019gtp,pardos2024mechanistic}, it broadly involves Mg coordinating a water molecule to initiate a nucleophilic attack on the terminal $\gamma$-phosphate group of GTP. This results in cleavage of the phosphate bond and formation of GDP and inorganic phosphate (P$_{i}$). During this hydrolytic process, the coordinated water molecule is cleaved into hydroxyl (OH) and proton (H$^{+}$) ions, which subsequently neutralize the free phosphate group, finalizing the biochemical reaction. The GTP hydrolysis mechanism then suggests four possible candidates for the RPM: GDP, P${i}$, OH, and Mg. Prior work has proposed a Mg$^{\text{\textbullet}+}$-P${i}$ radical pair \cite{LBuchachenko2014nn}, particularly in ATP-associated P${i}$ groups and protein kinase-mediated P${i}$ transfer via an ion-radical mechanism. However, studies suggest that while Mg$^{\text{\textbullet}+}$ radicals are chemically feasible and observed in specific non-biological contexts \cite{Jdrzkiewicz2022}, their formation in standard biological systems, such as microtubule polymerization, still requires experimental confirmation. Additionally, at the interatomic distance between Mg$^{\text{\textbullet}+}$ and P${i}$ on GTP-tubulin, the strong exchange interaction dominates, leading to rapid recombination or suppression of magnetic field sensitivity. Thus, at first approach, Mg and P${i}$ on the GTP appear as unlikely radical candidates for explaining observed magnetic field effects. 

Keeping this in mind, it must be noted that in the present work we focused on distance-independent Zeeman and hyperfine interactions, neglecting the distance-dependent exchange and dipolar interactions. Previous studies have suggested that dipolar and exchange interactions could neutralize each other at specific distances \cite{Efimova2008}. More recently, a triad radical model has been proposed as an effective method to mitigate the challenges posed by dipolar interactions \cite{Kattnig2021}. In practice, radical pairs are likely dynamic entities, frequently moving closer and farther apart. Such movement allow for variance in dipolar interactions which are averaged out over time, while maintaining sufficient distance most of the time to minimize exchange interactions. This dynamic nature ensures that radical pairs occasionally approach closely enough to permit chemical reactions while remaining sufficiently separated otherwise to preserve sensitivity to weak magnetic fields. Addressing this balance is a general challenge within the RPM, one which we acknowledge but do not explore in depth here. These dynamics may be further investigated in subsequent studies building upon recent insights into the flavin adenine dinucleotide biradical \cite{Sotoodehfar2024}. These further investigations may provide insight on potential radicals involved in the current microtubule system.

Beyond the GTP hydrolysis mechanism we need to consider other potential sources for radicals. The experimental biochemical assays contain the buffer components PIPES, EGTA, and glycerol. PIPES is an ethanesulfonic acid buffer employed to maintain a stable pH, while EGTA is a calcium-selective chelating agent commonly used in biological assays to control or suppress calcium ion concentration due to calcium's known depolymerizing effect on microtubules. Glycerol, a sugar alcohol included to stabilize microtubules, mimics the intracellular environment and contributes to maintaining microtubule integrity. Regarding radical formation, none of these buffer components are conventionally recognized to directly participate in, or significantly facilitate, radical formation under typical experimental conditions. 

To clarify the underlying chemistry in the RPM model, theoretical approaches could include quantum chemical modeling of radical formation pathways, particularly focusing on the stability and reactivity of Mg or OH radicals in the tubulin microenvironment. Experimentally, electron paramagnetic resonance spectroscopy could be employed to detect and characterize short-lived radical intermediates during microtubule assembly, providing direct evidence of their existence and magnetic properties. Prior research has proposed a tryptophan–superoxide radical pair affecting microtubule polymerization \cite{ZadehHaghighi2022MT} to explain HMF effects on microtubules \cite{Wang2008}. While the enzymatic GTPase process of tubulin, which involves the hydrolysis of GTP, is not known to directly produce superoxide, and superoxide is not specifically introduced into our biochemical assay. However, the potential for indirect radical generation, particularly via interactions with dissolved oxygen or other reactive species, cannot be entirely excluded and merits further investigation.

A broader objective is to develop a comprehensive model that fully explains the observed Mg isotope and magnetic field effects, including their implications for weak, low-frequency magnetic fields \cite{Wu2018} in biological systems. While our current framework captures key aspects of isotope-dependent radical pair dynamics in microtubules, further refinement is needed to integrate findings from \textit{in vivo} experiments, where additional biochemical complexity and environmental factors may influence these effects. This is of key importance to the broader applicability of the observed results. For example, disruptions in microtubule dynamics are closely linked to neurodegenerative diseases like Alzheimer’s Disease (AD) \cite{Congdon2018}. Notably, elevated brain Mg has been shown to exert synaptoprotective effects in AD models \cite{Li2013}, while weak magnetic fields are shown to influence learning and memory \cite{zhang2021long}. As our findings show a weak magnetic field influence on microtubule polymerization via an Mg sensitive RPM, this suggests external magnetic fields or isotopic variations may affect neuronal stability via a quantum-sensing mechanism. Investigating these quantum effects may open new avenues for understanding AD progression and developing novel therapeutic strategies.

In conclusion, the results presented here show weak magnetic field and magnesium isotope effects on microtubule polymerization dynamics that are consistent with a theoretical description of radical pair dynamics. The general RPM presented herein accurately explains the isotope effects in microtubule assembly, despite the underlying chemistry remaining unclear. This demonstrates the broad applicability of the RPM for understanding weak magnetic field effects and isotope effects in biology. Further characterization of the underlying chemistry and radical pairs involved in the microtubule system has the potential to bridge structural biology, biophysics, and quantum biology, with implications for neurobiology, bioengineering, and medical applications.

\section*{Methods}

\subsection*{\texorpdfstring{Enriched Mg isotopes}{Enriched Mg2+ Isotopes}}
Stable isotopes of \texorpdfstring{$^{25}$Mg and $^{26}$Mg}{25Mg2+ and 26Mg2+} were purchased in Mg oxide (MgO) form from BuyIsotope (Neonest AB, Solna, Sweden) at $>$ 99.38\% enrichment levels of isotope to ensure all effects observed are attributed to specific isotopes and not due to impurities. Natural abundance \texorpdfstring{$^{\text{Nat}}$MgO}{NatMgO} ($>$ 99.99\%) was purchased from Sigma-Aldrich (529699). MgO was converted to Mg chloride, MgCl\textsubscript{2}, through a reaction with hydrochloric acid, HCl, and used in tubulin polymerization assays.

\subsection*{Tubulin Polymerization Assays}
Tubulin polymerization assay kits with $>$ 99\% pure porcine tubulin were purchased from Cytoskeleton Inc. (BK006P). The manufacturer’s instructions were followed for the assays. General tubulin buffer from the kits was replaced with in-house made buffer using 80 mM PIPES pH 6.9 (6910-OP from Millipore), 0.5 mM EGTA (03777 from Sigma-Aldrich), and 2 mM MgCl$_2$ as described above. The eight central wells of a 96-well plate were plated with Mg isotope samples in alternating patterns on each plate with different alternating patterns for each run to allow each isotope equal placement across the plate and between the coils. Tubulin polymerization dynamics were measured via optical density (OD$_{355}$) using a Victor X4 microplate reader (Perkin Elmer) following the manufacturer’s instructions. Plates were read in the microplate reader for five minutes and then placed in an incubator set at 37 $^\circ$C containing a GSC 5-inch Helmholtz coil (Figure \ref{fig:figureS5}). After 50 minutes the plates were transferred back to the microplate reader and read for another five minutes. 

\subsection*{Field Measurements}
The magnetic field in each of the eight central wells of a 96-well plate were were measured with a hand-held digital high-precision Teslameter (Tunkia TD8620) with a GSC 5-inch Helmholtz coil powered on.  

\subsection*{Temperature Measurements}
Temperatures in each plate well were measured with a Traceable\textsuperscript{\textregistered} Lollipop\textsuperscript{TM} waterproof and shockproof thermometer (model 4371). 

\subsection*{Statistical Analysis}
Polymerization data for each replicate was normalized to the initial OD value for the replicate to account for slight timing differences in plating the tubulin solutions. All polymerization data was then normalized to the plate average OD for the last five minutes of the $^{26}$Mg replicates to account for plate-to-plate variations. Statistical comparisons for maximum tubulin polymerization were performed between final endpoints of conditions. Means and standard deviation of the mean (SEM) were calculated for each condition from their distributions. A two-way analysis of variance (ANOVA) was performed to analyze the effect of Mg isotope, presence of magnetic field, and interaction between isotope and field. A Tukey-Kramer post-hoc analysis was used to account for multiple comparisons.  To assess the effect of well-to-well temperature differences 

In all comparisons, a heteroscedastic two-tailed t-test with p-values of less than 0.05 was taken as significant. All statistical analysis was performed in MATLAB 2022a.

\subsection*{Tubulin Polymerization}

\begin{figure}[ht]
    \centering
    \includegraphics[width=0.8\textwidth]{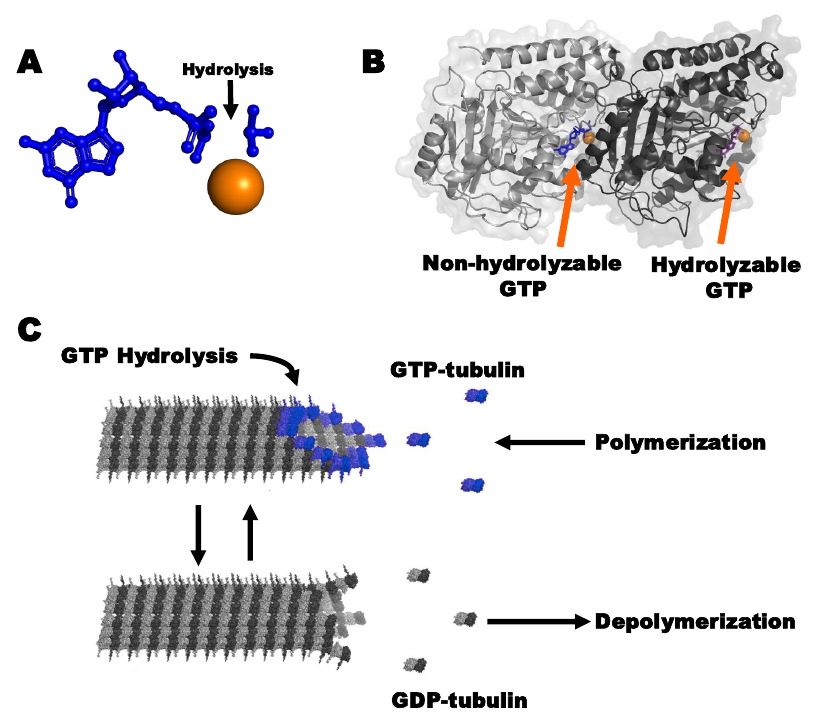} 
    \caption{\scriptsize\textbf{ Tubulin Polymerization} (A) GTP (blue) hydrolyzes in the presence of a Mg$^{2+}$ ion (orange sphere) to produce GDP. (B) The crystal structure of \(\alpha\)- (light grey),\(\beta\)- (dark grey) tubulin with GTP (blue) and GDP (violet) in the non-hydrolyzable and hydrolyzable sites. (C) $\alpha\beta$-tubulin containing GTP (blue tubulin) at the hydrolyzable site polymerizes by attaching to the growing end of a microtubules to form the left-handed, helical, chiral micro-tubule structure with a hollow center. As the microtubule grows tubulin hydrolyzes GTP with the aid of a Mg causing the tubulin protein to flex. If the GTP cap on the end of the growing microtubule disappears the flexed tubulin causes protofilaments to curve and separate from the microtubule causing the microtubule to depolymerize in catastrophe.  If additional GTP tubulin is added the microtubule is rescued and continues to grow. Modified image from \cite{Craddock2012} under a Creative Commons Attribution License.}   
    \label{fig:Figure5}
\end{figure}

Microtubules are dynamic protein polymers essential for eukaryotic cell structure, intracellular transport, and cellular organization \cite{Brouhard2018,Akhmanova2015,Kuo2021}. Microtubules self-assemble from the \(\alpha\), \(\beta\)-tubulin protein heterodimer through a Mg and guanosine triphosphate (GTP)-hydrolysis dependent mechanism (Figure \ref{fig:Figure5}A). Free tubulin subunits bind two GTP molecules, with one—the exchangeable GTP on \(\beta\)-tubulin—undergoing hydrolysis to guanosine diphosphate (GDP) in the presence of a magnesium ion (Mg$^{2+}$) \cite{LaFrance2022,Shim2024}, (Figure \ref{fig:Figure5}B).This conformational change, occurring after tubulin is incorporated into the microtubule, increases the dissociation rate of exposed tubulin subunits at the growing end, thereby regulating the polymerization and depolymerization dynamics of the microtubule (Figure \ref{fig:Figure5}C). As naturally occuring magnesium ($^{Nat}$Mg) exists in nature as three stable isotopes with varying abundances, \textsuperscript{24}Mg (79\%), \textsuperscript{25}Mg (10\%), and \textsuperscript{26}Mg (11\%), with only \textsuperscript{25}Mg possessing a nonzero nuclear spin (5/2+), it is uniquely suited to test spin-dependent isotope effects on GTP hydrolysis via measuring microtubule polymerization.

\subsection*{Model}
The chemical equation reads as follows:     

\begin{equation}
    \frac{d[MT(t)]}{dt}=k_p [Tu(t)]-k_d [MT(t)],
    \label{eq:mt}
\end{equation}
which yields $[MT(t)]= \frac{k_p [P]}{k_p+k_d} (1-e^{-(k_p+k_d)t})$.

The state of the radical pair is characterized by the spin density operator, which encapsulates the system's quantum state. The time evolution of the spin density matrix is influenced by a combination of coherent spin dynamics, chemical reactivity, and spin relaxation processes. This evolution is governed by the Liouville Master Equation, which provides the time dependence of the spin density operator for the radical pair system:

\begin{equation}
\frac{d\hat{\rho}(t)}{dt} = -\hat{\hat{L}}[\hat{\rho}(t)]. 
\end{equation}
Here, the Liouvillian superoperator is expressed as $\hat{\hat{L}} = i\hat{\hat{H}} + \hat{\hat{K}} + \hat{\hat{R}}$, where $\hat{\hat{H}}$, $\hat{\hat{K}}$, and $\hat{\hat{R}}$ represent the Hamiltonian superoperator, the chemical reaction superoperator, and the spin relaxation superoperator, respectively. These components collectively determine the dynamics of the spin density matrix. Further details about these terms can be found in the Supplementary Information. 

Beyond the conventional singlet and triplet states, radical pairs can also originate as F-pairs, which arise when two radicals form independently rather than through direct electron transfer between them. In this case, the initial state is a completely mixed state, represented as \(\frac{1}{4M} \hat{I}_{4M}\), where \(\hat{I}_{4M}\) is the identity matrix of dimension \(4M\) \cite{Kattnig2021}. $M$ is the nuclear spin multiplicity.

The Hamiltonian is expressed as follows:

\begin{equation}
    \hat{H}=\omega \hat{S}_{A_{z}}+a_A \mathbf{\hat{S}}_A.\mathbf{\hat{I}}_A+\omega \hat{S}_{B_{z}}+a_B \mathbf{\hat{S}}_B.\mathbf{\hat{I}}_B+a_{Mg} \mathbf{\hat{S}}_B.\mathbf{\hat{I}}_{Mg},
    \label{eq:zn-ham}
\end{equation}

Here, $\mathbf{\hat{S}}_A$ and $\mathbf{\hat{S}}_B$ represent the electron spin operators for the radical pairs labeled A and B, respectively. The Larmor frequency of the electrons, resulting from the Zeeman interaction, is denoted by $\omega$. Additionally, $\mathbf{\hat{I}}_A$ and $\mathbf{\hat{I}}_B$ correspond to the nuclear spin operator coupled with electron A, and B, respectively, while $\mathbf{\hat{I}}_{Mg}$ represents the nuclear spin operator of $^{25}$Mg coupled to electron B. The parameter $a_A$ and $a_B$ refers to the HFCC of nuclear spins coupled to electron A and B respectively. $a_{Mg}$ denotes the HFCC of $^{25}$Mg.

For spin-selective chemical reactions, we use the Haberkorn superoperator \cite{Steiner1989,Haberkorn1976}, which is given by the following equation:

\begin{equation}
\ket{S} = \frac{1}{\sqrt{2}}\left(\ket{\uparrow\downarrow} - \ket{\downarrow\uparrow}\right)
\end{equation}

\begin{equation}
\ket{T_0} = \frac{1}{\sqrt{2}}\left(\ket{\uparrow\downarrow} + \ket{\downarrow\uparrow}\right), \quad
\ket{T_{+1}} = \ket{\uparrow\uparrow}, \quad
\ket{T_{-1}} = \ket{\downarrow\downarrow}
\end{equation}

\begin{equation}
\hat{\hat{K}}=\frac{1}{2}k_S (\hat{P}^S\otimes I_{4M}+I_{4M}\otimes \hat{P}^S)+\frac{1}{2}k_T (\hat{P}^T\otimes I_{4M}+I_{4M}\otimes \hat{P}^T),
\end{equation}
where $\hat{P}^S$ and $\hat{P}^T$, respectively, are the singlet and triplet projection operators:
\begin{equation}
\frac{1}{M}\hat{P}^S =   \ket{S}\bra{S}\otimes \frac{1}{M}\hat{I}_M ,
\end{equation}
\begin{equation}
\frac{1}{3M}\hat{P}^T =  \frac{1}{3} \Big( \ket{T_0}\bra{T_0} + \ket{T_{+1}}\bra{T_{+1}} + \ket{T_{-1}}\bra{T_{-1}} \Big) \otimes \frac{1}{M}\hat{I}_M.
\end{equation}

Spin relaxation is modeled via random time-dependent local fields \cite{Kattnig2016,Player2020}, and the corresponding superoperator reads as follows:

\begin{equation}
\begin{aligned}
    \hat{\hat{R}} &= r_A \left[ \frac{3}{4} I_{4M} \otimes I_{4M} - \hat{S}_{A_{x}} \otimes (\hat{S}_{A_{x}})^T - \hat{S}_{A_{y}} \otimes (\hat{S}_{A_{y}})^T - \hat{S}_{A_{z}} \otimes (\hat{S}_{A_{z}})^T \right] \\
    &\quad + r_B \left[ \frac{3}{4} I_{4M} \otimes I_{4M} - \hat{S}_{B_{x}} \otimes (\hat{S}_{B_{x}})^T - \hat{S}_{B_{y}} \otimes (\hat{S}_{B_{y}})^T - \hat{S}_{B_{z}} \otimes (\hat{S}_{B_{z}})^T \right]
\end{aligned}
\end{equation}
where the symbols have the above-stated meanings. The ultimate fractional triplet yield for F-pairs, for time periods significantly exceeding the radical pair lifetime, is expressed as:

\begin{equation}
\Phi_T=k_T Tr[\hat{P}^T \hat{\hat{L}}^{-1}[\frac{\hat{I}_{4M}}{4M}]].
\label{eq:syr}
\end{equation}

\section*{Data availability}
The data used in this work can be found in Supplementary Information. 
\section*{Acknowledgment} 
We would like to thank Dr. Peter Hore and Dr. Jonathan Woodward for helpful and insightful discussions on potential mediators of a radical pair mechanism to explain our results, which contributed to the refinement of this work.  This research was undertaken, in part, thanks to funding to TJAC from the Canada Research Chairs Program and the University of Waterloo. CS was supported by the Natural Sciences and Engineering Research Council (NSERC) through its Discovery Grant program and the Alliance quantum consortia grant QuEnSI.
\section*{Author contributions}

TJAC, HZH and CS conceived the project; TJAC and RPS designed the experiments and provided experimental resources; HZH performed the theoretical modelling and calculations with help from TJAC and CS; CRS performed the experiments with help from RPS and TJAC; CRS and TJAC analyzed the experimental data; TJAC and HZH wrote the original draft with feedback from  CS; HZH, CRS, RPS, CS, TJAC reviewed and edited the final version.

\section*{Competing interests} The authors declare that they have no competing interests.

\bibliographystyle{unsrt}  
\bibliography{Ref}

\newpage

\section*{Supplementary Information}
\renewcommand{\thefigure}{S\arabic{figure}}
\setcounter{figure}{0}

\begin{figure}[ht]
    \centering
    \includegraphics[width=\textwidth]{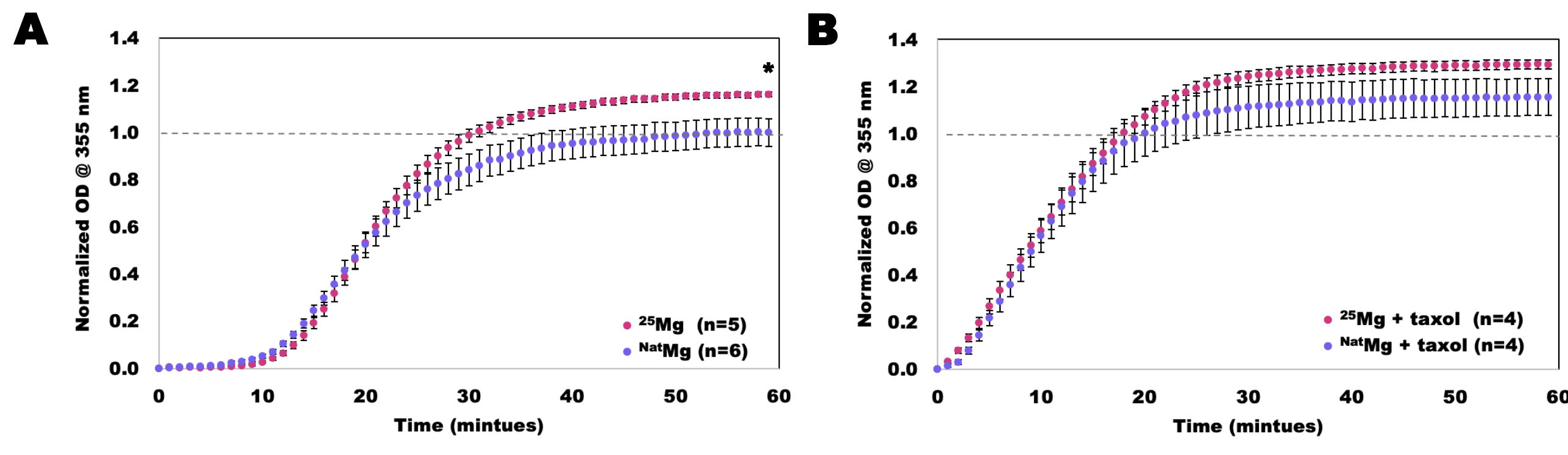} 
    \caption{\scriptsize\textbf{Tubulin Polymerization with In-House Buffers} Optical density (OD) curves over time showing tubulin polymerization in the presence of $^{25}$Mg and $^{Nat}$Mg in the absence of external magnetic field. (A) In the absence of the microtubule stabilizing agent taxol. (B) In the presence of 10 $\mu$M of the microtubule stabilizing agent taxol. OD values are in arbitrary units relative to the final values of tubulin polymerization with $^{Nat}$Mg  (dashed line). Points show mean ± standard deviation of the mean. * p = 0.033, difference from final $^{Nat}$Mg  without taxol by two-tailed heteroscedastic t-test.}
    \label{fig:figureS1}
\end{figure} 

\newpage

\begin{figure}[ht]
    \centering
    \includegraphics[width=1\textwidth]{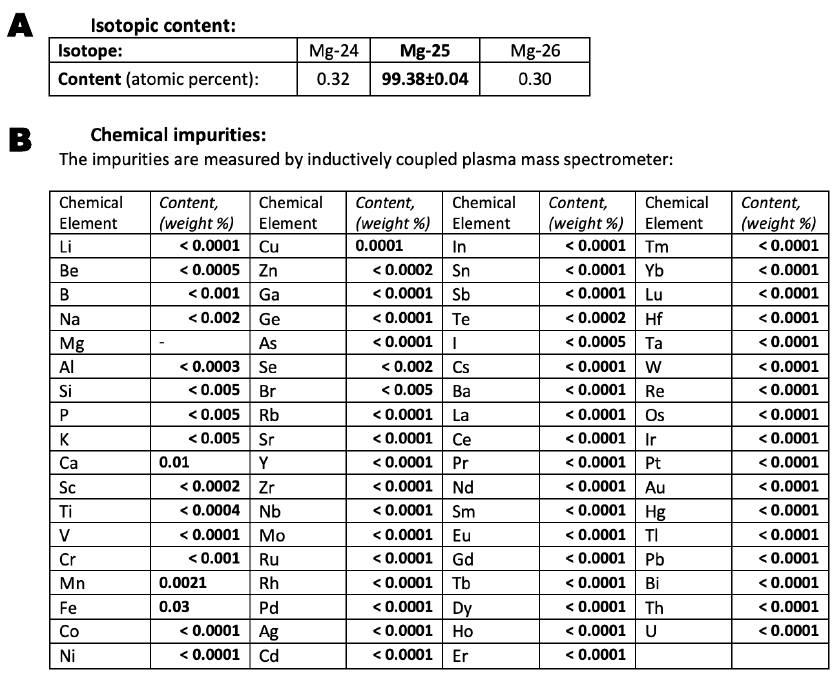} 
    \caption{\scriptsize\textbf{Composition of $^{25}$Mg } (A) Mg composition.  (B) Impurity composition. }
    \label{fig:figureS2}
\end{figure} 

\begin{figure}[ht]
    \centering
    \includegraphics[width=1\textwidth]{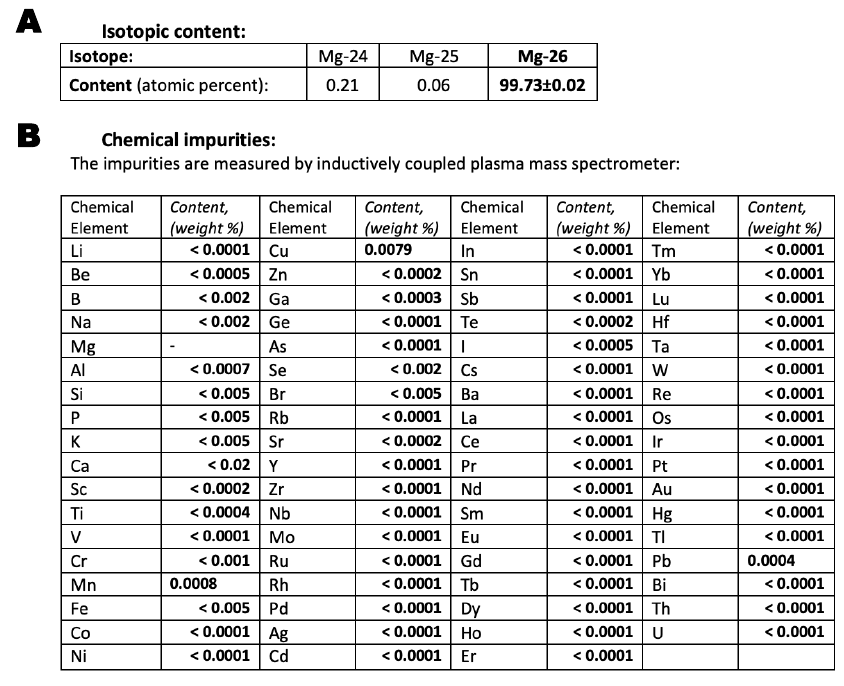} 
    \caption{\scriptsize\textbf{Composition of $^{26}$Mg } (A) Mg composition.  (B) Impurity composition. }
    \label{fig:figureS3}
\end{figure} 

\newpage

\begin{figure}[ht]
    \centering
    \includegraphics[width=1\textwidth]{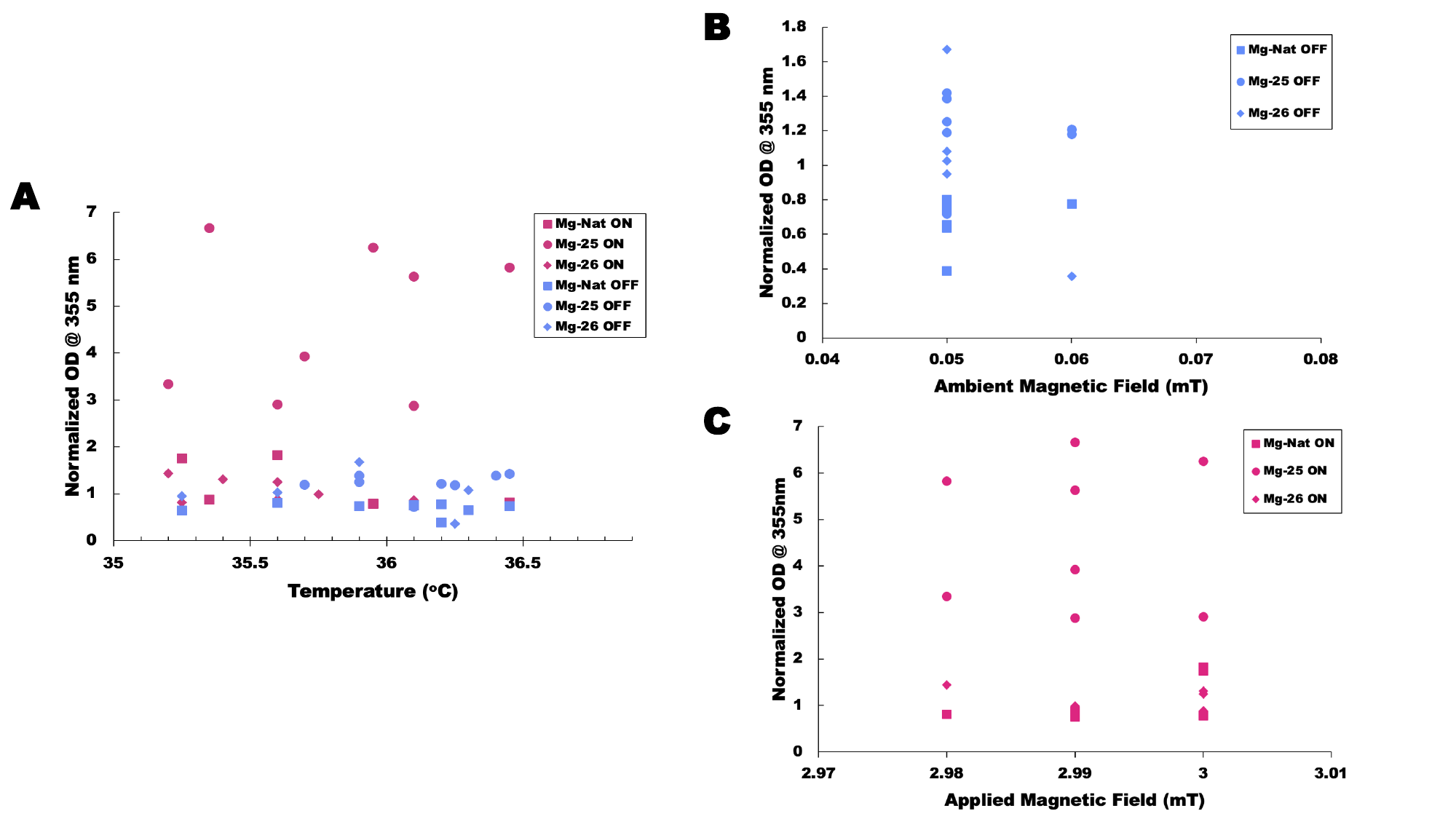} 
    \caption{\scriptsize\textbf{Isotope Effect on Tubulin Polymerization in the Presence of a 3 mT Magnetic Field.} (A) Scatter plot showing the final OD meaure of tubulin polymerization for $^{Nat}$Mg ($\blacksquare$), $^{25}$Mg ($\bullet$), and $^{26}$Mg ($\blacklozenge$) in the absence (blue) and presence (magenta) of the applied 2.99 \(\pm\) 0.02 mT magnetic field plotted against well-to-well measures of temperature. No significant correlations as determined by Pearson linear correlation, nor significant contribution to the observed effect as determined by N-way ANOVA analysis with a Tukey-Kramer posthoc test. (B) and (C) scatter plots showing the final OD meaure of tubulin polymerization for $^{Nat}$Mg ($\blacksquare$), $^{25}$Mg ($\bullet$), and $^{26}$Mg ($\blacklozenge$) in the absence (B; blue)  and presence (C; magenta) of the applied 2.99 \(\pm\) 0.02 mT magnetic field plotted against well-to-well measures of magnetic field. No significant correlations as determined by Pearson linear correlation. While it may be argued that we aren't capturing nT variations, the current spread is not likely to result in any significant correlations if these nT variations are accounted for.}
    \label{fig:figureS4}
\end{figure} 

\begin{figure}[ht]
    \centering
    \includegraphics[width=1\textwidth]{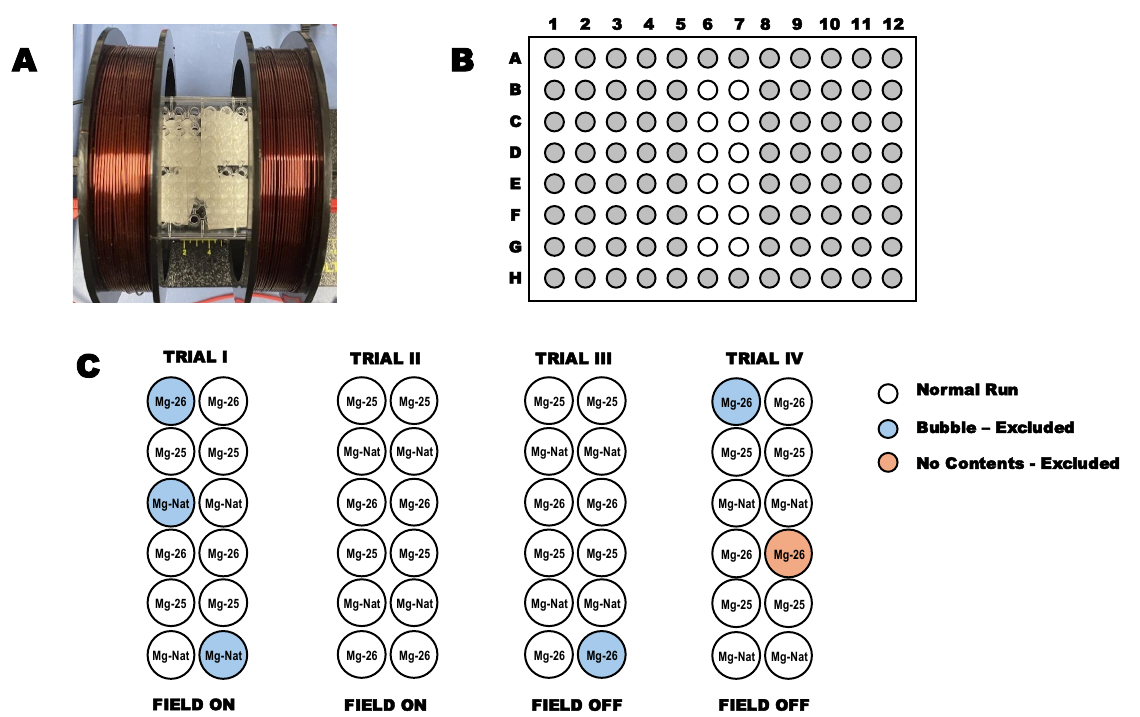} 
    \caption{\scriptsize\textbf{Experimental Setup and Magnesium Isotope Sample Arrangement.} (A) Experimental setup showing the coils and 96-well plate arrangement. 
(B) Schematic representation of the 96-well plate, with wells used in the experiment highlighted in white. 
(C) Layout indicating the placement of magnesium isotopes (\(^{Nat}\text{Mg}\), \(^{25}\text{Mg}\), \(^{26}\text{Mg}\)) across trials, noting the applied magnetic field (ON/OFF) condition.}
    \label{fig:figureS5}
\end{figure} 

\end{document}